\documentstyle[12pt,epsfig]{article}
\advance \topmargin by -\headheight
\advance \topmargin by -\headsep
\evensidemargin \oddsidemargin
\begin{document}
\pagestyle{plain}
\begin{titlepage}
\flushright{IHEP 2001-051}
\flushright{\today}
\vspace*{0.15cm}
\begin{center}
{\Large\bf
    Study of the $K^{-} \rightarrow \pi^{0} e^- \nu $ decay  
    }
\vspace*{0.15cm}

\vspace*{0.3cm}
{\bf  I.V.~Ajinenko, S.A.~Akimenko, G.A.~Akopdzhanov, K.S.~Belous, 
I.G.~Britvich, G.I.~Britvich,  A.P.~Filin, V.N.~Govorun,
A.V.~Inyakin, V.F.~Konstantinov, A.S.~Konstantinov,  
I.Y.~Korolkov, V.A.~Khmelnikov, V.M.~Leontiev, V.P.~Novikov,
V.F.~Obraztsov,  V.A.~Polyakov, V.I.~Romanovsky, V.M.~Ronjin, V.A.~Senko,
 N.A.~Shalanda, M.M.~Shapkin, V.I.~Shelikhov, N.E.~Smirnov, A.A.~Sokolov, 
  O.G.~Tchikilev, E.V.~Vlasov, O.P.~Yushchenko. }
  
\vskip 0.15cm
{\large\bf $Institute~for~High~Energy~Physics,~Protvino,~Russia$}

\vskip 0.35cm
{\bf V.N.~Bolotov, S.V.~Laptev, A.Yu.~Polyarush, V.E.~Postoev }
\vskip 0.15cm
{\large\bf $Institute~for~Nuclear~Research,~Moscow,~Russia$}
\vskip 0.35cm
{\bf  S.V.~Yaschenko, B.Zh.~Zalikhanov, V.Z.~Serdyuk. }
\vskip 0.15cm
{\large\bf $Joint~ Institute for Nuclear Research, ~Dubna,~Russia$}

\end{center}
\end{titlepage}
\newpage
\begin{center}
Abstract
\end{center}
 The  decay $K^{-} \rightarrow \pi^{0} e^- \nu$ has been 
 studied using in-flight decays detected with "ISTRA+" setup operating 
 at the 25 GeV negative secondary beam of the U-70 PS. About 130K events were
 used for the analysis. The $\lambda_{+}$ parameter of the vector formfactor 
 has been measured:
 $\lambda_{+}= 0.0293 \pm 0.0015(stat) \pm 0.002(syst)$. The limits on the
 possible tensor and scalar couplings have been derived: 

\begin{center}
 $f_{T}/f_{+}(0)=-0.044^{+0.059}_{-0.057}$ (stat) \\[0.3cm]  
 $f_{S}/f_{+}(0)=-0.020^{+0.025}_{-0.016}$ (stat) 
\end{center}

\newpage
\thispagestyle{empty}
\newpage
\raggedbottom
\sloppy

\section{ Introduction}
 
 The decay $K \rightarrow e \nu \pi^{0}$(K$_{e3}$) is known to be a 
 promising one
 to search for an  admixture of scalar (S) or tensor (T) interactions to the
 Standard Model (SM) V-A . This topic has been attracting 
 significant interest during resent years and moreover some previous experiments
 with charged and neutral kaon beams have
 reported indications for some anomalous S and T signals \cite{Akim1}, 
 \cite{KTeV}. On the other hand, resent KEK \cite{KEK1,KEK2}
 experiment with stopped $K^{+}$
 beam has reported negative results of the searches. 
 
 In 
 our analysis we present new search for S and T couplings based on the statistics 
 of about 130K Ke3 events.
  Another result of our study is the measurement of the V-A
 $f_{+}(t)$ formfactor slope $\lambda_{+}$.
  
\section{ Experimental setup}
The experiment is performed at the IHEP 70 GeV proton synchrotron U-70.
The experimental apparatus  "ISTRA+" is the result of the modification of
"ISTRA-M" \cite{ISTRA-M}, which, in turn, evolved from "ISTRA" that yielded 
important results on $\pi^{-}$ and $K^{-}$ decays in the late
1980's~\cite{ISTRA}.
 The setup is located at the 4A negative unseparated secondary beam. 
The beam momentum in the measurements is $\sim 25$ GeV with 
$\Delta p/p \sim 2 \%$. The admixture of $K^{-}$ in the beam is $\sim 3 \%$.
The beam intensity is $\sim 3 \cdot 10^{6}$ per 1.9 sec. U-70 spill.
A schematic view of the detector is shown in Fig.1. 
\begin{figure}
\epsfig{file=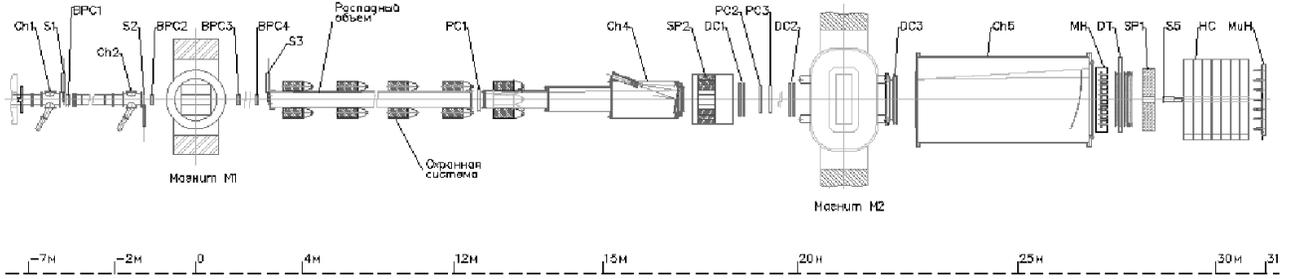,width=17cm}
\caption{ The layout of the   "ISTRA+" setup.}
\end{figure}
The momentum of the beam particle deflected by M$_{1}$ is 
measured by $BPC_{1}\div BPC_{4}$ PC's
with 1mm wire step, the kaon identification is done by 
$\check{C_{1}} \div \check{C_{3}}$ threshold $\check{C}$--counters. 
The 9 meter long vacuumated
decay volume is surrounded by 8 lead glass rings $LG_{1} \div LG_{8}$ used to
veto low energy photons. The same role is played by $SP_{2}$-- a 72--cell
lead glass calorimeter.
The decay products deflected in M2 with 1Tm field integral
are measured with $PC_{1} \div PC_{3}$-- 2mm step proportional chambers;
$DC_{1} \div DC_{3}$-- 1cm cell drift chambers and finally with 2cm diameter
drift  tubes    $DT_{1} \div DT_{4}$. 
A wide aperture threshold Cerenkov counter $\check{C_{4}}$ is 
filled with He and
used to  trigger the electrons. $SP_{1}$ is a 576-cell lead glass calorimeter,
followed by HC -- a scintillator-iron sampling hadron calorimeter, subdivided
into 7 longitudinal sections 7$\times$7 cells each. MH is a 
11$\times$11 cell scintillating hodoscope, used to solve the ambiguity 
for multitrack events and improve the time resolution of the 
tracking system, MuH  is a  7$\times$7 cell muon hodoscope. \\
The trigger is provided by $S_{1} \div S_{5}$ scintillation counters, 
$\check{C_{1}} \div \check{C_{3}}$ Cerenkov counters,
analog sum of amplitudes from the last dinodes of the $SP_1$ and 
is very loose:
 $T=S_{1} \cdot S_{2} \cdot S_{3} \cdot 
 \bar{S_{4}} \cdot \check{C_{1}} \cdot \bar{\check{C_{2}}} \cdot 
 \bar{\check{C_{3}}} \cdot 
 \bar{S_{5}} \cdot \Sigma(SP_{1})$,
here S4 is a scintillator counter with a hole to suppress beam halo ;
 $S5$ is a counter downstream  the setup at the beam focus;
$\Sigma(SP_{1})$ -- a requirement for the analog sum of amplitudes from 
$SP_1$ to be larger than $\sim$700 MeV -- a MIP signal. The last requirement 
surveys
to suppress the $K \rightarrow \mu \nu$ decay. Some complementary triggers:
$T_{e}=S_{1} \cdot S_{2}\cdot S_{3} \cdot 
 \bar S_{4} \cdot  \bar S_{5} \cdot \check{C_{4}}$ -- the electron
  trigger and prescaled "decay"
 trigger $T_{d}=S_{1} \cdot S_{2}\cdot S_{3} \cdot 
 \bar S_{4} \cdot \bar S_{5}$ were used to cross-check the efficiency 
 of the main one.

The main difference between "ISTRA-M" and "ISTRA+" is in the electronics and
DAQ: all the CAMAC based electronics was changed by IHEP developed MICC
\cite{MISS} ECL-based electronics. "ISTRA+" has now 12 MICC crates with ADC's,
TDC's
and latches. The DAQ, described in some details in \cite{DAQ} is based
on IHEP-developed VME master V-08 \cite{V-08}, which writes the MICC stream 
into standard
VME memory. Between the spills, the information is written into PC through
BIT-3 VME-PCI interface. The saturated event rate is $\sim 6500$ of 1~Kb 
events per 1.9 sec. spill.

\section{Event selection}
During physics runs in November--December 1999 and March--April 2001,
206M and 363M events were logged on DLT's. This information is supported
by about 100M MC events generated with Geant3 \cite{geant}. MC generation
includes realistic description of the setup: decay volume entrance windows,
track chambers windows, gas, sense wires and cathode structure,
Cerenkov counters mirrors and gas, shower generation in EM calorimeters etc.

The usual first step of the data processing is the EM calorimeter
calibration, using special runs with  10 GeV electrons; track system
alignment, HCAL and guard system calibration with muon beam runs.
The data processing starts with the beam particle reconstruction in 
$BPC_{1} \div BPC_{4}$, then the secondary tracks are looked for in 
$PC_{1} \div PC_{3}$ ; $DC_{1} \div DC_{3}$; $DT_{1} \div DT_{4}$
and events with one good negative track are selected.
The decay vertex is searched for, and a cut  is introduced on the
matching of incoming and decay track. The next step is to look for 
showers in $SP_{2}$ calorimeter. A method of shower parameters reconstruction 
based on the MC--generated patterns($\sim 2000\; 3 \times 3$ patterns) of showers 
is used. The matching of the charged track and a shower in $SP_{2}$ is done 
on the basis of the difference between the track extrapolation and the 
shower coordinates. The electron identification is done using E/P ratio --
of the energy of the shower associated with the track and the track momentum,
see Fig.2. 
\begin{figure}
\begin{center}
\epsfig{file=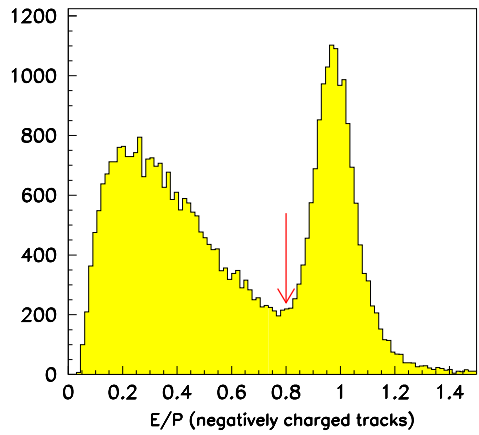,width=7cm}
\end{center}
\caption{ The E/p plot -- the ratio of the energy of the associated
cluster in ECAL to the momentum of the charged track. The arrow shows the
cut used for the electron separation.}
\end{figure}
 The selection of events with the two extra showers results in 
$M_{\gamma \gamma}$ spectrum shown in Fig.3.
\begin{figure}
\begin{center}
\epsfig{file=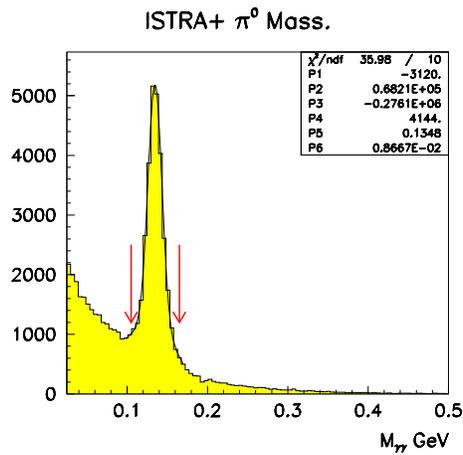,width=7cm}
\end{center}
\caption{ The $ \gamma \gamma$ mass spectrum for the events with
the identified electron and two extra showers.}
\end{figure}
The $\pi^{0}$ peak
has a mass of $M_{\pi0}=134.8 MeV$, and a  resolution of 8.6 MeV.
Another important variable for the $K^{-} \rightarrow e \nu \pi^{0}$ 
selection is the missing mass squared: $(P_{K}-P_{e}-P_{\pi^{0}})^{2}$,
where P are the corresponding four-momenta, see Fig.4. The cut is 
$\pm 0.01$ GeV$^{2}$.
\begin{figure}
\begin{center}
\epsfig{file=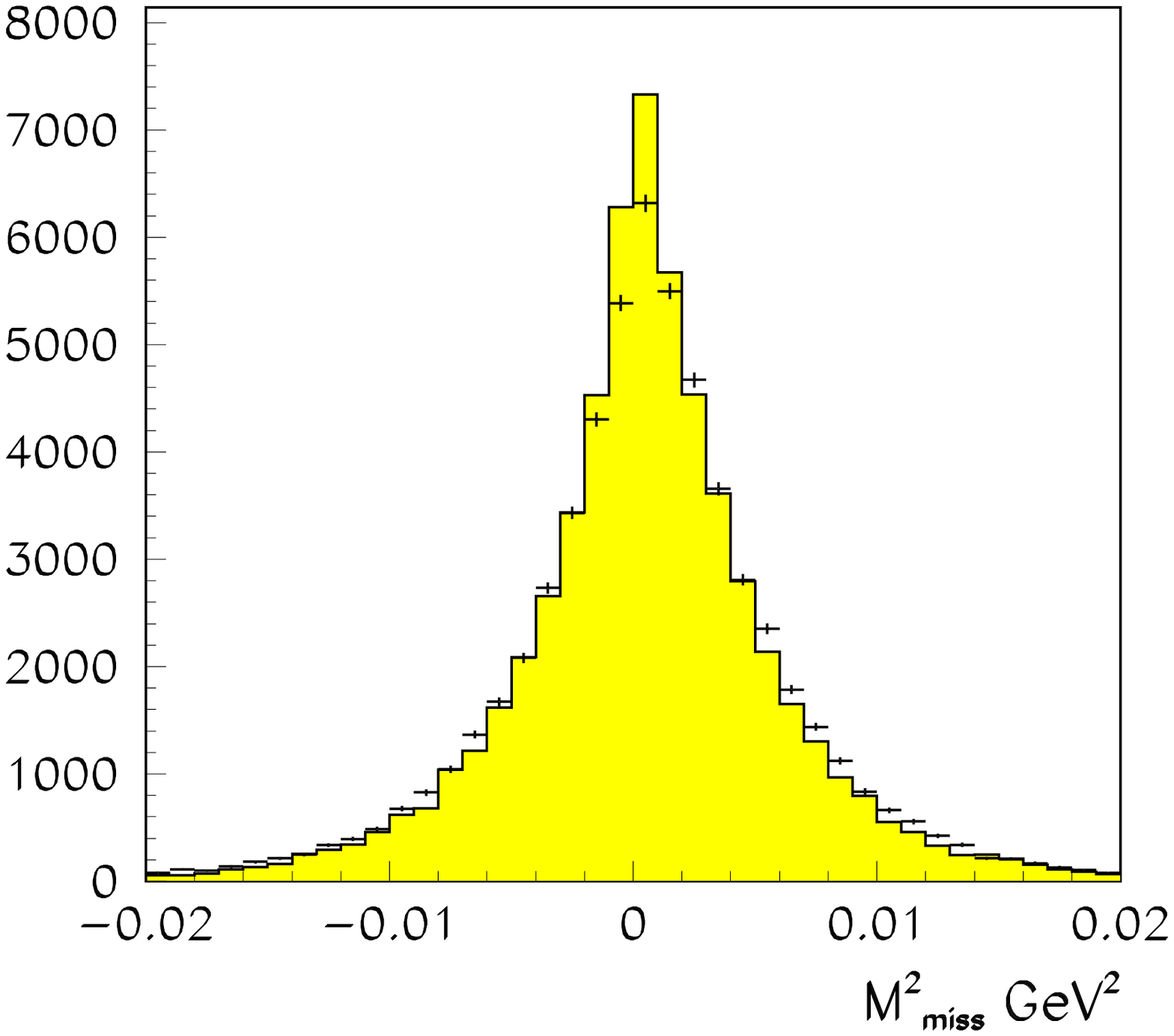,width=7cm}
\epsfig{file=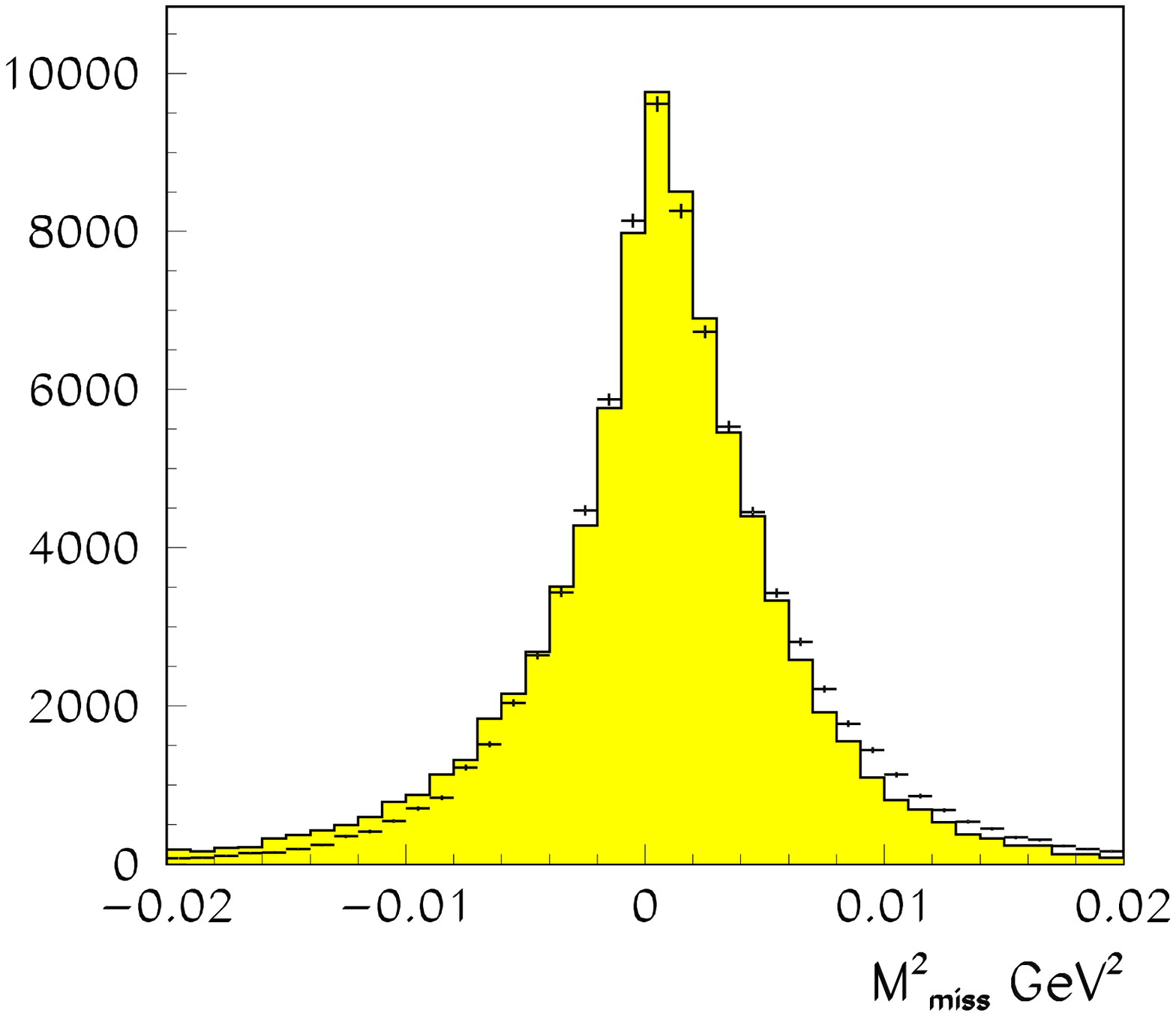,width=7cm}
\end{center}
\caption{The missing four-momentum squared $(P_{K}-P_{e}-P_{\pi^{0}})^{2}$
for the selected events for 1999 run (left) and 2001 run (right). The
points with errors are the data, the histogram -- MC. }
\end{figure} 
The further selection  is done by the requirement that the event passes 2C   
 $K \rightarrow e \nu \pi^{0}$ fit.
At the same time, similar  2C fit  $K \rightarrow
\pi^{-} \pi^0$ should fail.
The missing energy $E_{K}-E_{e}-E_{\pi^{0}}$ after this selection is shown 
in Fig.5
The peak at low $E_{miss}$ corresponds to
the remaining $K^{-} \rightarrow \pi^{-} \pi^{0}$ background. The corresponding
cut is $E_{miss}>1GeV$.
 The surviving background is estimated from MC to be less
than $3\%$.  
\begin{figure}
\begin{center}
\epsfig{file=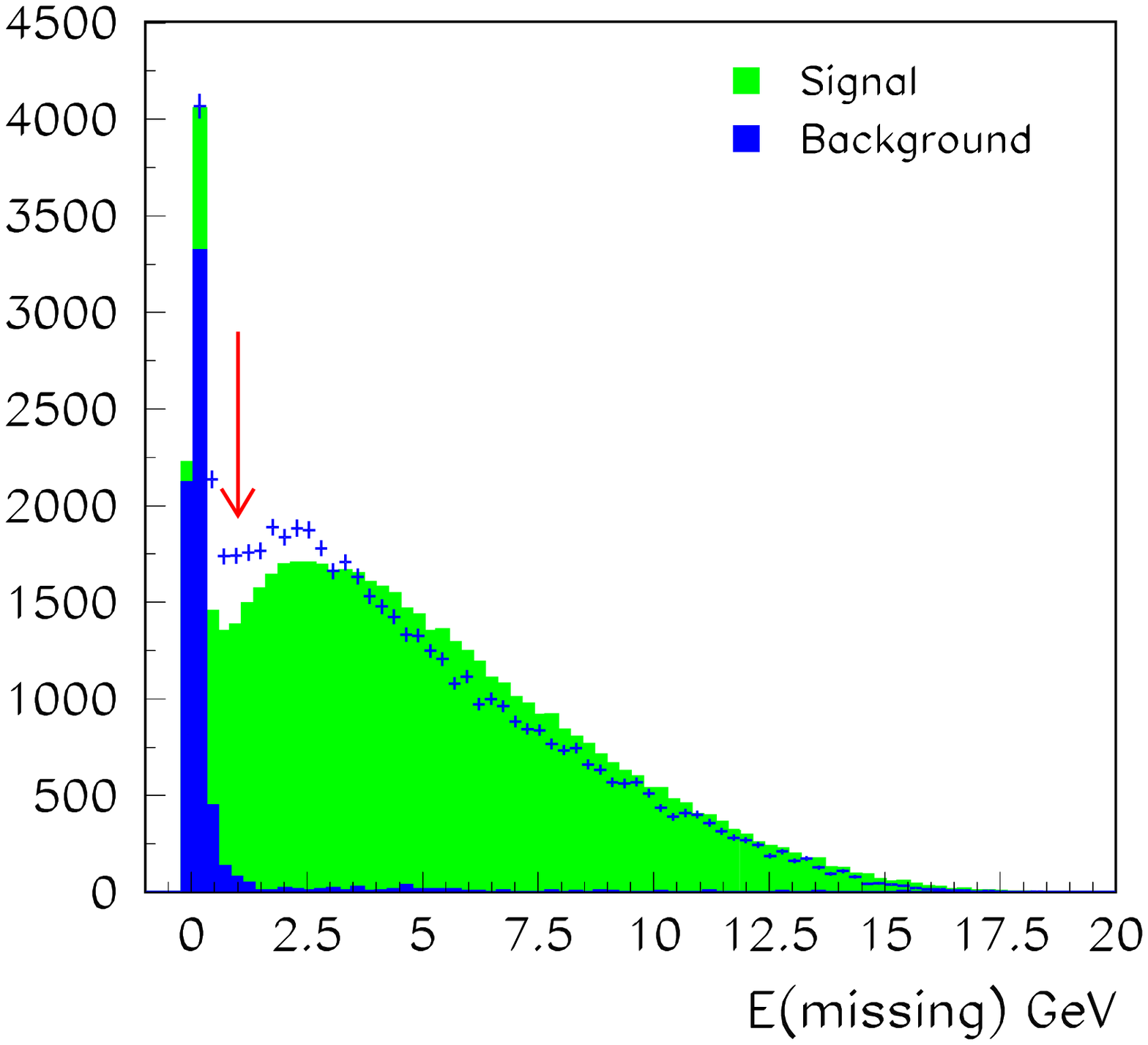,width=7cm}
\epsfig{file=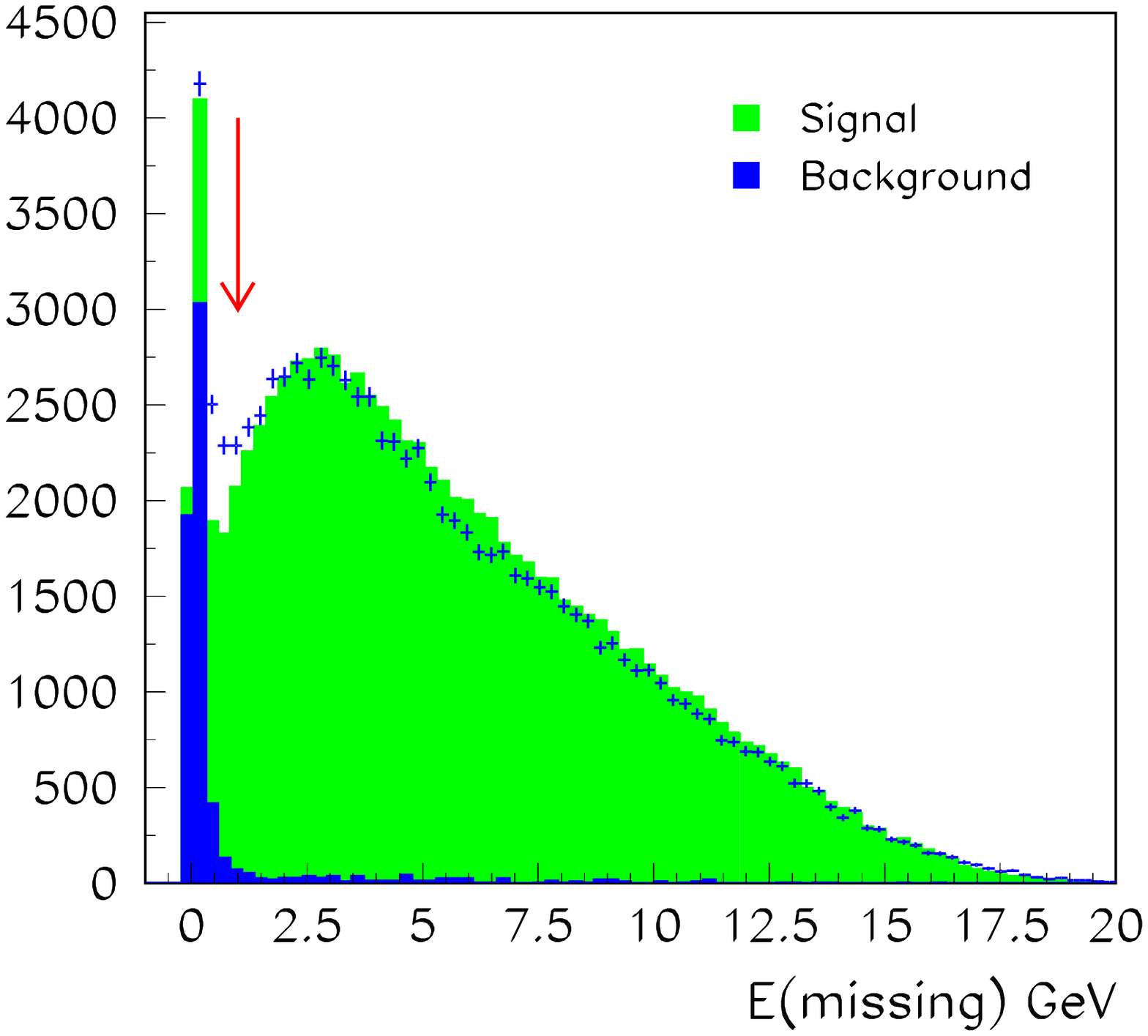,width=7cm}
\end{center}
\caption{The missing energy for the $e \pi^{0}$ events, for 1999 run (left)
and 2001 run (right). The points with errors are the data, the histograms -- MC.
The dark(blue) peak at zero value corresponds to the 
MC-predicted $K \rightarrow \pi^{-} \pi^{0}$ background. 
The arrow indicates the cut value.}
\end{figure}
 
The detailed data reduction information is shown in Table.1.
\begin{table}[bth]
\caption{ Event reduction statistics. The main steps are shown for 
the 1999 and 2001 runs .}                                 

\renewcommand{\arraystretch}{1.5}
\begin{center}
\begin{tabular}{|c|c|c|}
\hline
 Run  &     1999 &           2001  \\  
\hline
 $N_{events}$ on tapes &  206.544.909 & 363.002.105 \\
\hline
Beam track reconstructed  &     159.459.629=77$\%$ & 268.564.958 =74 $\%$ \\
\hline
 One secondary track found &     81.166.929=41$\%$ & 134.227.095 =37$\%$  \\                    
\hline
 Written to DST  &    70.015.610 =34$\%$ &107.215.783 =30 $\%$ \\
\hline \hline
 $e^{-}$ identified &     1.300.958 & 1.998.719 \\
\hline
 2 showers reconstructed &  252.177 & 361.621 \\
\hline
$\pi^{0}$ identified   & 186.850 & 251.489 \\
\hline
 $|M_{miss}^{2}|<0.01$ & 96.652 & 144.642  \\
\hline
$K \rightarrow \pi^{-} \pi^{0}$ rejected &79.660 &117.566 \\
\hline
$K \rightarrow e \nu \pi^{0}$ accepted & 65.208 & 97.585 \\
\hline
$E_{miss} >$ 1 GeV & 54.009 & 79.248 \\
\hline
\end{tabular}
\end{center}
\end{table}
\renewcommand{\arraystretch}{1.0}

\section{ Analysis}
The event selection described in the previous section results in selected
54K events in 1999 data and 79K events in 2001 data. The distribution of
the events over the Dalitz plot is shown in Fig.6. The  variables 
$y=2E_{e}/M_{K}$ and $z=2E_{\pi}/M_{K}$, where $E_{e}$, $E_{\pi}$ are the
energies of the electron and $\pi^{0}$ in the kaon c.m.s are used. The
background events, as MC shows, occupy the peripheral part of the plot. 

\begin{figure}
\begin{center}
\epsfig{file=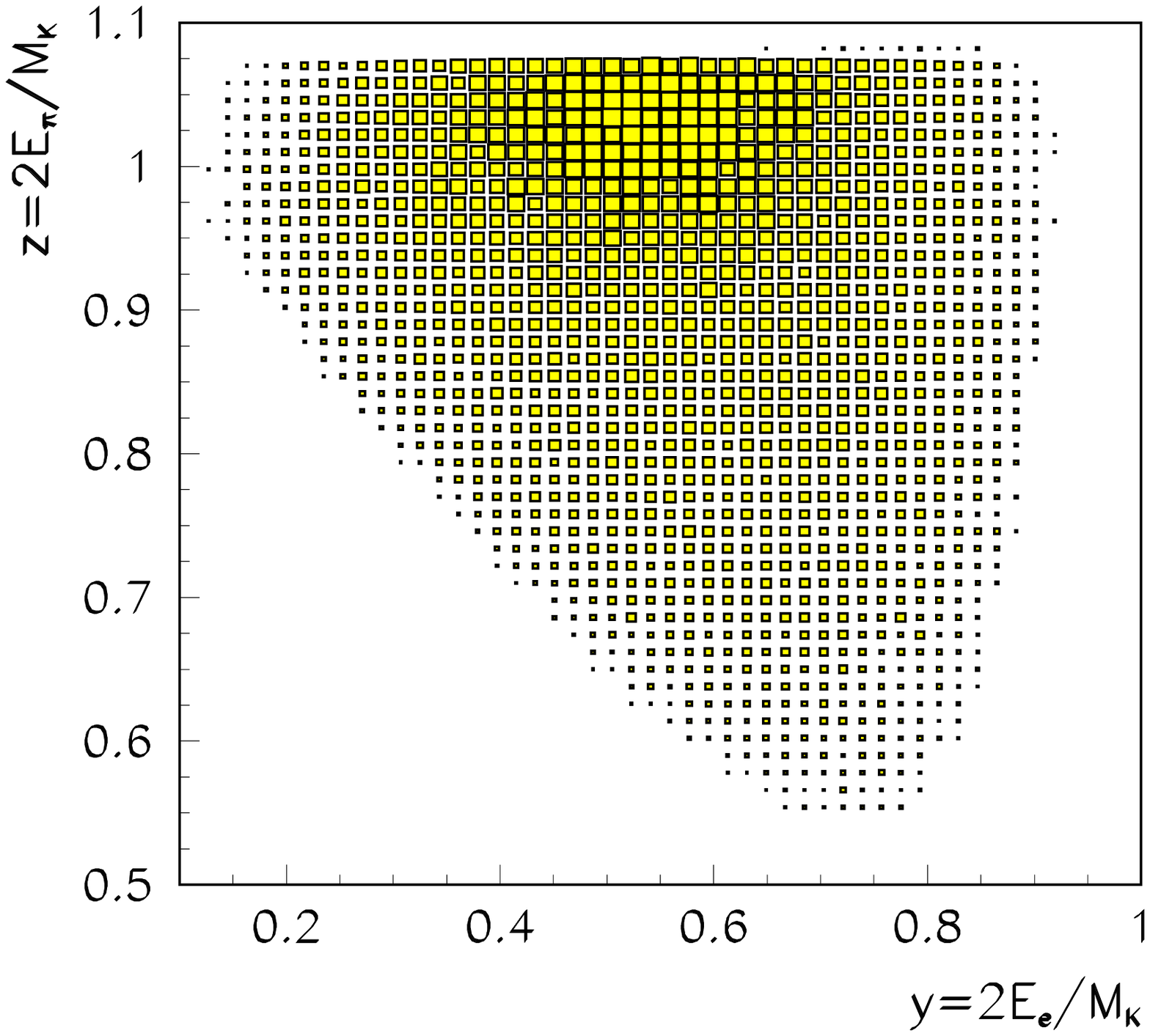,width=7cm}
\epsfig{file=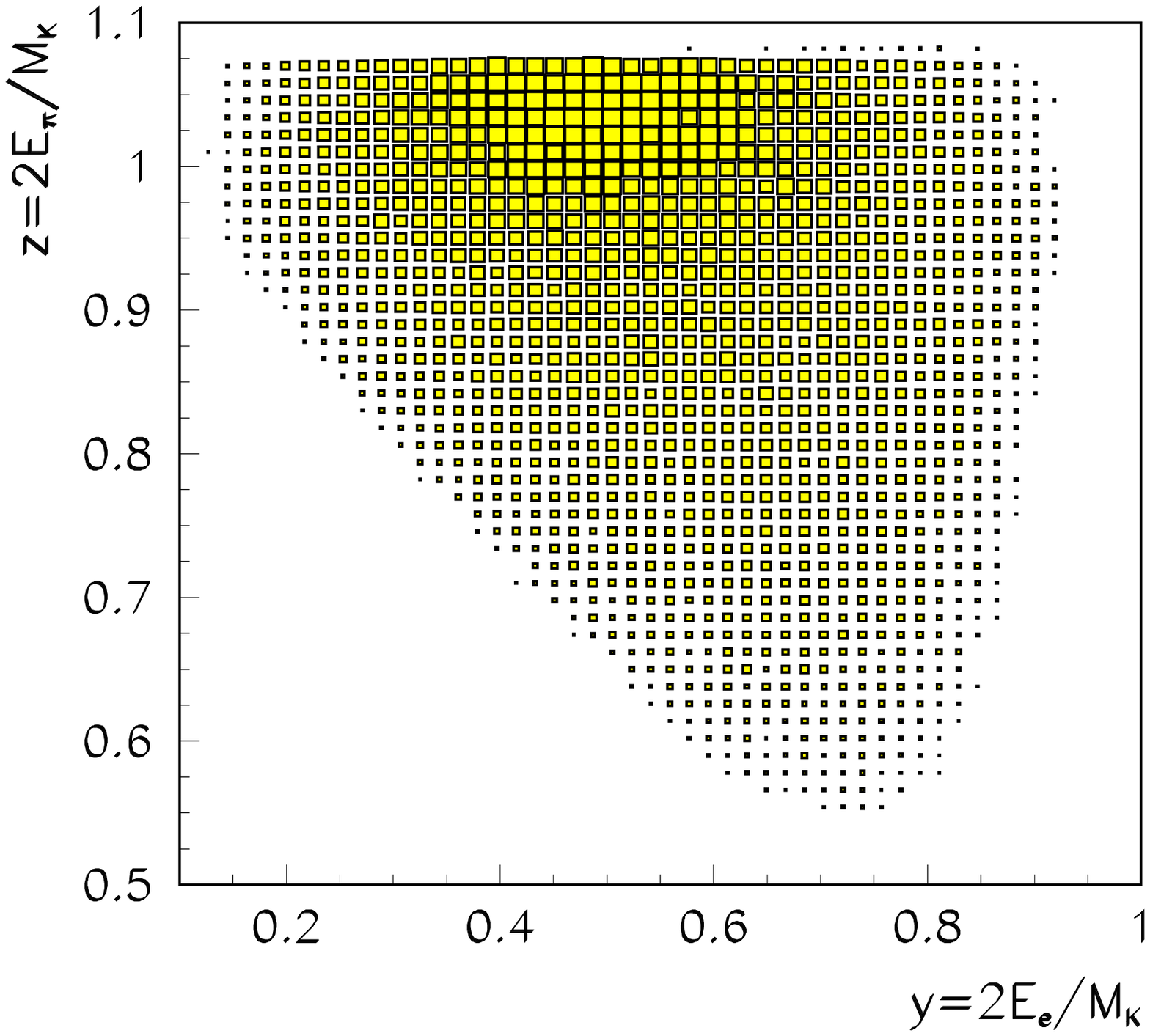,width=7cm}
\end{center}
\caption{ Dalitz plots $(y=2E_{e}/M_{K} ;z=2E_{\pi^{0}}/M_{K})$ for
the selected $K \rightarrow e \nu \pi^{0}$ events after the 2-C fit.
Left -- 1999 statistics, Right -- 2001 statistics.}
\end{figure}
The most general Lorentz invariant form of the matrix element for the 
decay $K^{-} \rightarrow l^{-} \nu \pi^{0}$ is  \cite{Steiner}:
\begin{equation}
M= \frac{G_{F}sin\theta_{C}}{\sqrt{2}} \bar u(p_{\nu}) (1+ \gamma^{5})
[m_{K}f_{S} -
\frac{1}{2}[(P_{K}+P_{\pi})_{\alpha}f_{+}+
(P_{K}-P_{\pi})_{\alpha}f_{-}]\gamma^{\alpha} + i \frac{f_{T}}{m_{K}}
\sigma_{\alpha \beta}P^{\alpha}_{K}P^{\beta}_{\pi}]v(p_{l})
\end{equation}
It consists of scalar, vector and tensor terms. $f_{S},f_{T}, f_{\pm}$
are the functions of $t= (P_{K}-P_{\pi})^{2}$. In the Standard Model (SM)
the W-boson exchange leads to the pure vector term. The "induced" 
scalar and/or tensor terms, due to EW radiative corrections are negligible
small, i.e the nonzero scalar/tensor form factors indicate a physics
beyond SM. 

The term in the vector part, proportional to $f_{-}$ is reduced(using Dirac
equation) to a scalar formfactor. In the same way, the tensor term is reduced to
a mixture of a scalar and a vector formfactors. The redefined $f_{+}$(V), 
$F_{S}$(S) and the corresponding Dalitz plot
density in the kaon rest frame($\rho(E_{\pi},E_{l})$) are \cite{Chizov}:
\begin{eqnarray}
 V & = & f_{+}+(m_{l}/m_{K})f_{T} \nonumber \\ 
S & = & f_{S} +(m_{l}/2m_{K})f_{-}+
\left( 1+\frac{m_{l}^{2}}{2m_{K}^{2}}-\frac{2E_{e}}{m_{K}}
-\frac{E_{\pi}}{m_{K}}\right) f_{T} \nonumber \\ 
\rho (E_{\pi},E_{l}) & \sim & A \cdot |V|^{2}+B \cdot Re(V^{*}S)+C \cdot |S|^{2} \\
A & = & m_{K}(2E_{l}E_{\nu}-m_{K} \Delta E_{\pi})-  
m_{l}^{2}(E_{\nu}-\frac{1}{4} \Delta E_{\pi}) \nonumber \\
B & = & m_{l}m_{K}(2E_{\nu}-\Delta E_{\pi}) \nonumber \\
C & = & m_{K}^{2} \Delta E_{\pi} \nonumber \\
\Delta E_{\pi} & = & E_{\pi}^{max}-E_{\pi} \nonumber
\end{eqnarray}
In case of Ke3 decay one can neglect the terms proportional to 
$m_{l}$; $m_{l}^{2}$. Then, assuming  linear dependence of $f_{+}$
on t: $f_{+}(t)=f_{+}(0)(1+\lambda_{+}t/m_{\pi}^{2})$ and real constants 
 $f_{S}$, $f_{T}$  we get:  
\begin{eqnarray}
\rho (E_{\pi},E_{l}) & \sim &
 m_{K}(2E_{l}E_{\nu}-m_{K} \Delta E_{\pi})\cdot
  (1+\lambda_{+}t/m_{\pi}^{2})^{2} \nonumber \\
 & + & m_{K}^{2} \Delta E_{\pi}\cdot \left( \frac{f_{S}}{f_{+}(0)} +
\left( 1-\frac{2E_{e}}{m_{K}}
-\frac{E_{\pi}}{m_{K}}\right) \frac{f_{T}}{f_{+}(0)}\right)^{2}     
\end{eqnarray}
The procedure for the experimental extraction of the parameters
$ \lambda_{+}$, $f_{S}$, $f_{T}$ starts from the subtraction of the MC
estimated background from the Dalitz plots of Fig.6. The background 
normalization was determined by the ratio of the real and generated
$K^{-} \rightarrow \pi^{-} \pi^{0}$ events. Then the Dalitz plots were
subdivided into 20 $\times$ 20 cells.
The background subtracted  distribution of the numbers of  events in the cells
(i,j) over Dalitz plots, for example, in the case of 
simultaneous extraction of $\lambda_{+}$ and $\frac{f_{S}}{f_{+}(0)}$,
was fitted with the function:
\begin{eqnarray}
\rho (i,j)\sim W_{1}(i,j)+W_{2}(i,j) \cdot \lambda_{+}+
W_{3}(i,j) \cdot \lambda_{+}^{2}+ W_{4}(i,j) 
\cdot \left( \frac{f_{S}}{f_{+}(0)}\right)^{2}      
\end{eqnarray} 
Here $W_{l}$ are MC-generated  functions, which are build up as follows:
the MC events are generated with constant density over the Dalitz plot and
reconstructed with the same program as for the real events.  Each event
carries the weight w determined by the corresponding term in formula 3,  
calculated using the MC-generated values for y and z.  
The radiative corrections according to \cite{grinb} were taken into account.
Then $W_{l}$ is constructed by summing up the weights of the events in
the corresponding Dalitz plot cell. This procedure allows to avoid the
systematics errors due to the "migration" of the events over the Dalitz plot
because of the finite experimental resolution.
\section{Results}
The results of the fit are summarized in Table.2.
The combination of the two runs is done by the simultaneous fit.
The first line corresponds to  pure V-A SM fit. In the second line
the tensor  and in the third the scalar terms are  added into the fit.
 All the errors presented are from the "MINOS" procedure of the "MINUIT"
program \cite{Minuit} and are larger than the Gaussian ones.
At present, we estimate
an additional systematics error in $\lambda_{+}$ to be $\pm 0.002$. The 
estimate is done
by comparing two  runs, which differ a lot in amount of matter in the
beamline and detector configuration and by varying cuts, cell size during the
fit of the Dalitz plots etc.

\renewcommand{\arraystretch}{2.2}

\begin{table}[bth]
\caption{ Results of the fit.}                                 

\begin{center}
\begin{tabular}{|c|ccc|}
\hline
    & 1999  & 2001 & 1999+2001 \\ \hline\hline
    
  $\lambda_+$ &  $~~0.0271^{+0.0023}_{-0.0023}$ & 
                         $~~0.0310^{+0.0019}_{-0.0019}$  &
                         $~~0.0293^{+0.0015}_{-0.0015}$ \\ \hline		 
   $\lambda_+$ &  $~~0.0270^{+0.0023}_{-0.0023}$ & 
                         $~~0.0310^{+0.0019}_{-0.0019}$  &
                         $~~0.0293^{+0.0015}_{-0.0015}$ \\    
  $f_T = F_T/f_+(0)$       &  $-0.0388^{+0.0878}_{-0.0848}$ &
                         $-0.0487^{+0.0806}_{-0.0754}$ &
                         $-0.0445^{+0.0597}_{-0.0574}$   \\ \hline			  
   $\lambda_+$ &  $~~0.0268^{+0.0024}_{-0.0027}$ & 
                         $~~0.0304^{+0.0022}_{-0.0024}$  &
                         $~~0.0289^{+0.0017}_{-0.0018}$ \\    
  $f_S = F_S/f_+(0)$       &  $-0.0139^{+0.0338}_{-0.0259}$ &
                         $-0.0225^{+0.0355}_{-0.0187}$ &
                         $-0.0197^{+0.0252}_{-0.0163}$   \\ \hline\hline

  $\chi^2/\mbox{ndf}$ &  1.7  & 1.3 & 1.5 \\  \hline
  $N_{\mbox{bins}}$ & 225    & 228  &  \\ \hline\hline
\end{tabular}
\end{center}
\end{table}

\renewcommand{\arraystretch}{1.0}

 The comparison of our results with the
most recent $K^{\pm}$ data \cite{Akim1,KEK1,KEK2} shows that our
statistics, at present, is the highest in the world and the statistical errors
somewhat smaller than in \cite{Akim1,KEK1} and comparable with \cite{KEK2}.
We do not confirm the observation of
a significant $f_{S}$ and $f_{T}$ in \cite{Akim1}. Our data is in a good 
agreement with \cite{KEK1,KEK2} and with the theoretical calculations for
$\lambda_{+}$:  $\lambda_{+}=0.031$ \cite{DAFNE}, done in the context of
the chiral perturbation theory.  
 
\section{Summary and conclusions}
The $K^{-}_{e3}$ decay has been studied using in-flight decays of 25 GeV 
$K^{-}$, detected by "ISTRA+" magnetic spectrometer. Due to the high
statistics, adequate resolution of the detector and good sensitivity over
all the Dalitz plot space, the measurement errors are significantly reduced
as compared with the previous measurements. 
 The $\lambda_{+}$ parameter of the vector formfactor 
 has been measured to be: 
\begin{center} 
 $\lambda_{+}= 0.0293 \pm 0.0015(stat) \pm 0.002(syst)$.
\end{center}
   The limits on the
 possible tensor and scalar couplings have been derived: 
\begin{center} 
 $f_{T}/f_{+}(0)=-0.044^{+0.059}_{-0.057}; $ \\[3mm]
 $f_{S}/f_{+}(0)=-0.020^{+0.025}_{-0.016} $ 
\end{center}


\begin{thebibliography}{99}
\bibitem{Akim1} S.~A.~Akimenko  et al., Phys. Lett. {\bf B259}(1991)225.
\bibitem{KTeV} R.~J.~Tesarek, hep-ex/9903069, 1999.
\bibitem{KEK1} S.~Shimizu et al., Phys. Lett. {\bf B495}(2000)33.
\bibitem{KEK2} A.S.~Levchenko et al., hep-ex/0111048(2001).
\bibitem{ISTRA-M} V.~N.~Bolotov  et al., 
 "Experimental Setup ISTRA-M to study
 rare decays of charged light mesons", IHEP 95-111, Protvino, 1995. 
\bibitem{ISTRA}V.~N.~Bolotov  et al., Yad.Fiz., {\bf v44}(1986)108,117; 
 {\bf v45}(1986)1652. 
\bibitem{MISS} Yu.B.Bushnin et al., IHEP 88-47, Serpukhov, 1988;
V.A.Medovikov  et al., IHEP 99-60, Protvino, 1999;
A.N.Isaev  et al., Prib.Tekh.Exp. {\bf N3}(2000)41;
M.M.Vasiliev  et al., Prib.Tekh.Eksp. {\bf N5}(2000)45. 
\bibitem{DAQ} A.Filin  et al., "The Linux Based Distributed Data 
Acquisition System for the ISTRA+ Experiment" Proceedings of the CHEP2001
Conference, Bejing, 2001.
\bibitem{V-08} V.N.~Govorun et al., IHEP preprint in preparation.
\bibitem{geant} R. Brun et al., CERN-DD/EE/84-1.
\bibitem{Steiner} H.~Steiner et al., Phys.Lett. {\bf B36}(1971)521.
\bibitem{Chizov} M.V.Chizhov hep-ph/9511287(1995).
\bibitem{grinb} E.S.Grinberg, Phys. Rev. 162 (1967), 1570.
\bibitem{Minuit} F.James, M.Roos, CERN D506,1989.
\bibitem{DAFNE} The DA$\Phi$NE Physics Handbook, v1, p.39, Frascati,1992
and references therein.
\end{thebibliography}
\end{document}